\documentclass[aps,pre,reprint,superscriptaddress]{revtex4-1}

\usepackage[dvips]{graphicx}
\usepackage{color}

\begin{document}

\preprint{accepted to PRE}

\title{Broadening of Cyclotron Resonance Conditions in the
  Relativistic Interaction of an Intense Laser with Overdense Plasmas}


\author{Takayoshi Sano}
\email{sano@ile.osaka-u.ac.jp}
\affiliation{Institute of Laser Engineering, Osaka University, Suita,
  Osaka 565-0871, Japan} 

\author{Yuki Tanaka}
\affiliation{Institute of Laser Engineering, Osaka University, Suita,
  Osaka 565-0871, Japan} 

\author{Natsumi Iwata}
\affiliation{Institute of Laser Engineering, Osaka University, Suita,
  Osaka 565-0871, Japan} 

\author{Masayasu Hata}
\affiliation{Institute of Laser Engineering, Osaka University, Suita,
  Osaka 565-0871, Japan} 

\author{Kunioki Mima}
\affiliation{Institute of Laser Engineering, Osaka University, Suita,
  Osaka 565-0871, Japan} 
\affiliation{The Graduate School for the Creation of New Photonics
  Industries, Hamamatsu, Shizuoka 431-1202, Japan} 

\author{Masakatsu Murakami}
\affiliation{Institute of Laser Engineering, Osaka University, Suita,
  Osaka 565-0871, Japan} 

\author{Yasuhiko Sentoku}
\affiliation{Institute of Laser Engineering, Osaka University, Suita,
  Osaka 565-0871, Japan} 


\date{\today}

\begin{abstract}
The interaction of dense plasmas with an intense laser under a strong
external magnetic field has been investigated.      
When the cyclotron frequency for the ambient magnetic field is higher
than the laser frequency, the laser's electromagnetic field is
converted to the whistler mode that propagates along the field line.     
Because of the nature of the whistler wave, the laser light penetrates
into dense plasmas with no cutoff density, and produces superthermal
electrons through cyclotron resonance. 
It is found that the cyclotron resonance absorption occurs effectively
under the broadened conditions, or a wider range of the external
field, which is caused by the presence of relativistic electrons
accelerated by the laser field.   
The upper limit of the ambient field for the resonance increases in
proportion to the square root of the relativistic laser intensity.  
The propagation of a large-amplitude whistler wave could raise the
possibility for plasma heating and particle acceleration deep inside
dense plasmas. 
\end{abstract}


\maketitle


\section{Introduction}

Remarkable progress has been made in the generation of an extremely
strong magnetic field over kilo Tesla by using high power lasers
\cite{yoneda12,fujioka13,santos15,korneev15,goyon17}.  
Laser plasma interaction in such a field condition is now attracting
much attention \cite{yang15,feng16,luan16,gong17}.    
Existence of a strong field affects the laser-generated high energy
density plasmas by microscopic energy transport and turbulence
\cite{braginskii65,schekochihin04} as well as by macroscopic
hydrodyanmics and instabilities
\cite{sano13,albertazzi14,nagatomo15,matsuo17}.
To understand those processes is quite important for the various
applications such as particle acceleration \cite{arefiev16}, inertial
confinement fusion (ICF)
\cite{hohenberger12,perkins13,wang15,fujioka16}, and laboratory laser
astrophysics \cite{remington06}.    

In this work, we focus on the propagation of a whistler wave in
overdense plasmas.   
The electron cyclotron frequency $\omega_{ce}$ for a kilo-Tesla field
becomes comparable to the laser frequency $\omega_0$.   
Here the critical field strength, of which the cyclotron frequency
equals $\omega_0$, is defined as $B_c \equiv m_e \omega_0 / e \approx$
10 ($\lambda_0 /1 \mu$m)$^{-1}$ kT, where $m_e$ is the electron rest
mass, $e$ is the elementary charge, and $\lambda_0$ is the laser
wavelength.  
The laser light can enter overdense plasmas as a whistler wave when
$\omega_{ce} > \omega_0$, because the cutoff frequency disappears. 
Furthermore, the whistler wave has another unique aspect that is the
cyclotron resonance with electrons \cite{chen84}.
These features have a crucial meaning in laser plasma interaction,
since the direct interaction between a high intensity laser and
overdense plasmas could bring a new mechanism of efficient plasma
heating and particle acceleration.  

Electron acceleration associated with whistler waves have been
studied in planetary magnetosphere plasmas
\cite{omura07,summers07,horne08}.  
It should be noted that the strong field ($\omega_{ce} > \omega_0$) and
overdense ($\omega_{pe} > \omega_0$, where $\omega_{pe}$ is the plasma
frequency) situations we are considering here is also appropriate in
the planetary plasmas. 
For example, in the Jovian magnetosphere, $\omega_{pe} / \omega_0
\sim 50$ and $\omega_{ce} / \omega_0 \sim 10$ for a kHz whistler mode
\cite{horne08}. 
However, the essential difference would be the whistler wave's
amplitude when compared with the external field strength. 
For ultra-intense laser cases, the relativistic effects by the
large-amplitude wave could make substantial changes in the wave
propagation and energy conversion processes.  

In this paper, we investigate the cyclotron resonance caused by an
external magnetic field while taking into account the effects of
relativistic electrons. 
Although the topic has been widely studied \cite{liu04,belyaev08,arefiev15},
our particular goal is to reveal the influence of the laser intensity
on the resonance conditions for the field strength.
Section \ref{sec:2} presents analytical consideration for the
derivation of the resonance conditions in the interaction between an
intense laser and magnetized overdense plasmas. 
The validity of the predicted conditions is confirmed numerically
by a series of one-dimensional (1D) Particle-in-Cell (PIC) simulations
of Sec. \ref{sec:3}. 
In Sec. \ref{sec:4}, we will discuss the applications of the resonant
properties of a ultra-intense laser under a strong magnetic field.  

\section{Derivation of Resonance Conditions \label{sec:2}}

First, let us consider the cyclotron resonance condition for a
right-hand (R) circularly polarized (CP) laser, that is, a
large-amplitude whistler wave with a frequency $\omega_0$ and a
wavenumber $k_0$.  
The relativistic Doppler-shifted cyclotron resonance condition is
given by $\omega_0 - k v_{\parallel} = {\omega_{ce}}/{\gamma}$, where
$k$ and $v_{\parallel}$ are the wavenumber and electron velocity along 
an external magnetic field $B_{\rm ext}$ of which the cyclotron
frequency $\omega_{ce} = e B_{\rm ext} / m_e$, and $\gamma$ is the
electron's Lorentz factor.    
Concentrating on a case where the wavenumber is identical to that of
the incident whistler wave, $k = k_0$, the condition is rewritten as 
\begin{equation} 
\omega_0 \left( 1 - \frac{\beta_{\parallel}}{\beta_{\phi}} \right) 
= \frac{\omega_{ce}}{\gamma} \;,
\label{eq:condition}
\end{equation}
where $\beta_{\ast} \equiv v_{\ast} / c$ stands for the velocities
normalized by the speed of light $c$, and $v_{\phi} = \omega_0 / k_0$
is the phase velocity of the whistler mode.
This relation contains two important factors which affect the
resonance condition, namely the Doppler shift and relativistic
effects.   

The second term of left-hand side of Eq. (\ref{eq:condition})
originates from the Doppler effect.   
The phase velocity $v_{\phi}$ of whistler waves in the electron
density $n_e$ can be much smaller than $c$, 
\begin{equation}
\beta_{\phi}^2
\approx \left( 
\frac{\omega_{pe}^2/\omega_0^2}{\omega_{ce}/\omega_0-1} 
\right)^{-1} \ll 1 \;,
\label{eq:vph}
\end{equation}
in strongly magnetized overdense plasmas ($\omega_0 < \omega_{ce} \ll
\omega_{pe}$).  
Here, $\omega_{pe} = [ e^2 n_e / (\varepsilon_0 m_e) ]^{1/2}$ is the
plasma frequency and $\varepsilon_0$ is the vacuum permittivity. 
The relativistic modification of dispersion relation for the whistler
branch will only change by a small factor \cite{sotnikov05}, so that this  
non-relativistic formula for $v_{\phi}$ may not affect the qualitative
interpretation in the following discussion. 

The resonant field strength is obtained as $\omega_{ce} / \omega_0
\approx 1 + (\beta_{\parallel} \omega_{pe} / \omega_0 )^{2/3}$ from
Eqs. (\ref{eq:condition}) and (\ref{eq:vph}) in the non-relativistic
limit of $\gamma \approx 1$.  
The solution exists only for the electrons traveling in the opposite
direction of the whistler wave, $v_{\parallel} v_\phi < 0$. 
This relation determines the resonant field strength as a function
of the parallel velocity.
Assuming $v_{\parallel}$ is of the order of the thermal velocity
$v_{\rm th}$, it gives 
\begin{equation}
\tilde{B}_{\rm res} \approx 1 + 
\beta_{\rm th}^{2/3}
\tilde{n}_e^{1/3} \;,
\label{eq:doppler}
\end{equation}
where $\tilde{B}_{\ast} \equiv B_{\ast} / B_c$, $\tilde{n}_e \equiv
n_e / n_c$, and $n_c = \varepsilon_0 m_e \omega_0^2 / e^2$ is the
critical density.
This condition indicates a small upper shift of the
resonant field $B_{\rm res}$ from the critical value $B_c$, of which
the cyclotron frequency $\omega_{ce} = \omega_0$.
For the underdense cases, $n_e \ll n_c$, there is no gap from $B_c$
because of the fast phase velocity $v_{\phi} \sim c$.   

The relativistic effect in Eq. (\ref{eq:condition}) can
alter the resonance condition much more significantly. 
After the laser interaction, the electrons in dense plasmas moves
mostly perpendicular to the external field by gyration.
Ignoring the parallel velocity $v_{\parallel} \ll c$, the
resonance condition for $\gamma$ is given by $\gamma_{\rm res}
\approx \omega_{ce} / \omega_0$, or
\begin{equation}
\gamma_{\rm res} 
\approx \tilde{B}_{\rm ext} \;.
\label{eq:gres}
\end{equation}
On the other hand, the quiver energy accelerated by the electric field
$E_0$ of the CP laser is about $\gamma_q = (1 + a_0^2)^{1/2}$, where
$a_0 = e E_0 / (m_e c \omega_0^2)$ is the normalized vector potential
of the laser light and the intensity is given by $I_0 = \varepsilon_0
c E_0^2$. 

Only the laser-accelerated electrons that have acquired a larger
energy than $\gamma_{\rm res}$ can get a chance for the resonance, so
that the required condition is expressed as $\gamma_q \gtrsim
\gamma_{\rm res}$.
Therefore, together with Eq. (\ref{eq:doppler}), the upper limit of
the resonant strength for $B_{\rm ext}$ is summarized as   
\begin{equation}
  \tilde{B}_{\rm upper} \sim
\max \left\{
1 + \beta_{\rm th}^{2/3} \tilde{n}_e^{1/3} \;,\;
\left( 1 + a_0^2 \right)^{1/2} \right\} \;.
\label{eq:bres}
\end{equation}
In the non-relativistic limit, the cyclotron resonance occurs only at
the specific condition of $\tilde{B}_{\rm ext} = 1$.
However, it turns out that the resonant condition has a wider range
when including the relativistic effect, which is stimulated by the
large-amplitude whistler wave.

\section{Validation by Numerical Simulations \label{sec:3}}

\begin{figure*}
\includegraphics[scale=1.0,clip]{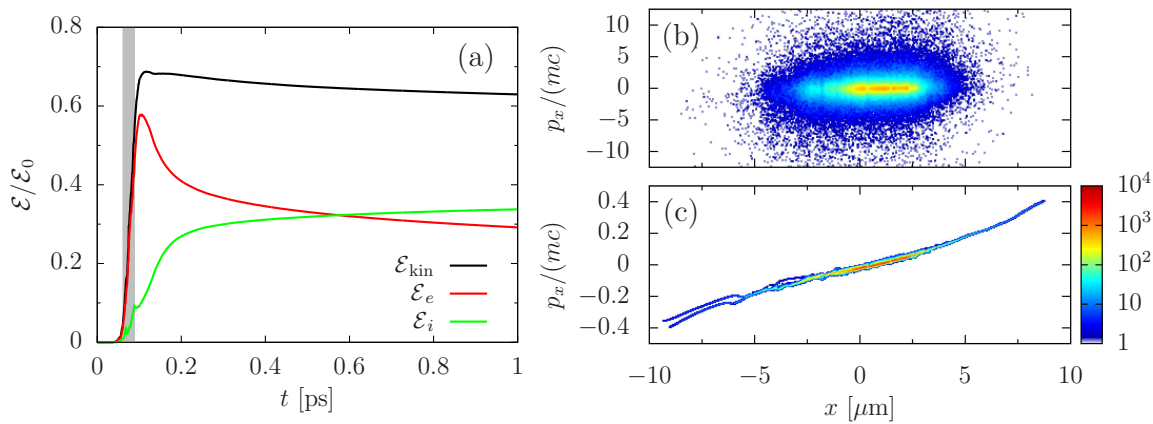}%
\caption{
(Color online)
(a) 
Temporal evolution of plasma kinetic energies in the fiducial model
for the electrons $\mathcal{E}_e$ (red), ions $\mathcal{E}_i$ (green),
and total $\mathcal{E}_{\rm kin} = \mathcal{E}_e + \mathcal{E}_i$ (black),
which are normalized by the incident laser energy $\mathcal{E}_0$. 
The gray area denotes the interaction duration of the laser with the
plasma foil.  
(b,c)
Phase diagram of the position $x$ and momentum in the $x$ direction
$p_x$ for (b) the electrons and (c) ions, which are taken at the end 
of calculation $t = 1$ ps.  
The color shows the particle number.
The initial location of the target foil is from $x = 0$ to 1 $\mu$m.
\label{fig:1}}
\end{figure*}

Next, we have tested this predicted condition, Eq. (\ref{eq:bres}), by
performing a series of 1D collisionless PIC simulations
\cite{matsumoto93}.  
In this numerical experiment, the interaction between a femto-second
laser and a solid-density foil is examined. 
A thin hydrogenic target with 1 $\mu$m thickness is irradiated by a 
Gaussian-shape laser with 30 fs duration at half maximum amplitude.
The electron number density of the target is set to be constant
$\tilde{n}_e = 151$ ($n_e = 2.63 \times 10^{23}$ cm$^{-3}$).  
For simplicity, the foil is in contact directly with the vacuum
without any preplasma.  

The laser is assumed to be a RCP light traveling in the $x$
direction. 
The wavelength is 0.8 $\mu$m, so that the critical magnetic field of
this pulse is $B_c = 13.4$ kT.  
The main pulse reaches the target at $t \approx 60$ fs, and the
simulation runs until 1 ps.
The laser intensity is varied from weakly relativistic of $I_0 =
10^{18}$ ($a_0 = 0.484$) to strongly relativistic $10^{21}$ W/cm$^2$
($15.3$).
Collisionless approximations would be appropriate in this regime
\cite{kemp06}, while the collisional absorption is dominant at the 
much lower intensity \cite{price95,luan16}. 
A uniform external magnetic field $B_{\rm ext}$ is applied along the
laser propagation direction, which is constant in time throughout the
computation in 1D situation.   

For the PIC scheme, the Debye length is adopted as the size of the
cell, where the initial electron temperature is assumed to be $T_e =
0.5$ keV unless otherwise mentioned.  
The particle number per cell for both electrons and ions is initially
not less than 100 in all runs. 
The computational domain is sufficiently large, so that no particle
and no electromagnetic field can reach the boundaries during the
simulations. 

Figure \ref{fig:1}(a) shows the typical time history of the kinetic
energies of electrons $\mathcal{E}_e$ and ions $\mathcal{E}_i$
converted from the incident laser energy $\mathcal{E}_0$.
Each particle has the kinetic energy $\epsilon$, and $\mathcal{E}_e$
($\mathcal{E}_i$) is the sum of all electrons (ions) in the
simulation box.
The initial parameters in this fiducial model are $B_{\rm ext} = 50.1$ kT
($\tilde{B}_{\rm ext} = 3.74$) and $I_0 = 10^{21}$ W/cm$^2$.
During the interaction with the main pulse ($t \sim 60$-$90$ fs), the
electrons gain a large amount of kinetic energy, and then it is
transferred to the ions gradually through the expansion process 
seen in Figs. \ref{fig:1}(b) and \ref{fig:1}(c).
The total energy converted to electrons and ions reaches almost 70\%
of $\mathcal{E}_0$, and which is one order of magnitude higher than
that in the unmagnetized case ($\ll$ 10\%). 

\begin{figure*}
\includegraphics[scale=1.0,clip]{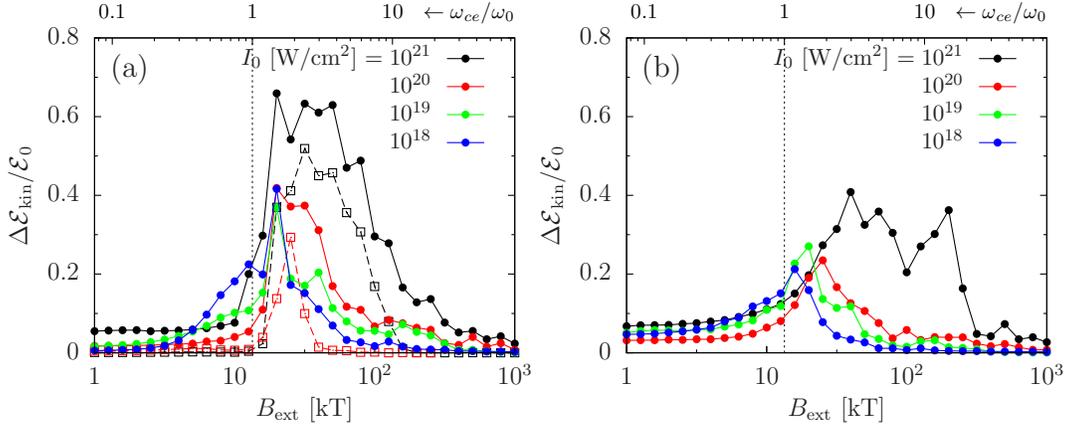}%
\caption{
(Color online)
(a)
Energy absorption evaluated by the gain of the total plasma
energy $\Delta \mathcal{E}_{\rm kin}$ as a function of the external 
field strength $B_{\rm ext}$ for the CP laser cases of $I_0 = 10^{21}$
(black), 10$^{20}$ (red), 10$^{19}$ (green), and 10$^{18}$ W/cm$^2$
(blue). 
The critical field strength $B_c$ is indicated by the dotted line.
For the purpose of comparison, the results of the ion immobile runs
for $I_0 = 10^{21}$ and $10^{20}$ W/cm$^2$ are also shown by the open
squares. 
(b) 
The same figure as panel (a) for the LP laser cases.
\label{fig:2}}
\end{figure*}

\begin{figure}
\includegraphics[scale=1.0,clip]{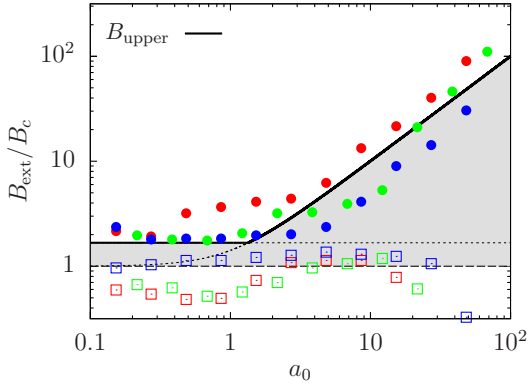}%
\caption{
(Color online)
Upper limits of the external field strength $B_{\rm ext}$ (filled
circles) for the range where the absorption rate is enhanced over a
threshold value, which are shown as a function of
the normalized intensity of the incident laser $a_0$. 
The different colors indicate different model assumptions.
The red and green marks are for the cases with the RCP and LP laser,
respectively.
The blue ones are with the RCP laser, but the ions are assumed to
be immobile.
The solid curve shows the theoretical prediction for
$\widetilde{B}_{\rm upper}$ given by Eq. (\ref{eq:bres}) assuming
$\widetilde{B}_{\rm res} = 1.7$.
The lower limits obtained numerically are also plotted by the open
squares.
\label{fig:6}}
\end{figure}

Figure \ref{fig:2}(a) depicts the energy absorption, which is the
conversion efficiency from the laser to plasmas, evaluated at the end
of calculations in various models with the different field strength
$B_{\rm ext}$ and laser intensity $I_0$.  
It is obvious that huge enhancement of the energy absorption takes
place near the critical strength $B_{\rm ext} \sim B_c$.
If $\tilde{B}_{\rm ext} \lesssim 0.1$, the external field makes little
difference in the laser plasma interaction.
Note that, for the lower intensity cases (e.g., $I_0 = 10^{18}$
W/cm$^2$), the peak is not exactly at $B_c$, but always
shifted to a slightly larger strength due to the Doppler effect
discussed above. 
The deviation estimated from Eq. (\ref{eq:doppler}) is $\tilde{B}_{\rm
  res} \sim 1.7$ for this model, which is in good agreement with the
numerical results.  

When the laser intensity increases, the resonant absorption occurs
with the wider range of $B_{\rm ext}$. 
For the cases with $I_0 = 10^{21}$ W/cm$^2$, large absorption is
achieved even when $\tilde{B}_{\rm ext} \sim 10$.   
In order to check the ion contribution in the absorption process,
simulations with immobile ions were performed which exhibits a 
similar trend to the cases with mobile ions [see open marks in
  Fig. \ref{fig:2}(a)]. 
Therefore, we can conclude that the high absorption with the wider range
of $B_{\rm ext}$ is caused by the cyclotron resonance of electrons
with the relativistic effect inferred by Eq.\,(\ref{eq:bres}).  

It should be emphasized that the same conclusion is expected for the
models with a linearly polarized (LP) laser [see Fig. \ref{fig:2}(b)]. 
In dense plasmas, a LP light splits into left-hand (L) CP and RCP
lights. 
Although the L wave part is reflected at the L cutoff ($\tilde{n}_e  =
1 + \tilde{B}_{\rm ext}$), 
the whistler part can propagate into and interact directly with the
plasma. 
In fact, when the laser polarity in the fiducial model switches
to linear one, the acceleration by cyclotron resonance is still
observed with the similar manner. 
But the fractional absorption becomes 33\%, which is just half of the
original CP model (63\%). 

To confirm the broadening of the cyclotron resonance,
we can distinctly observe in Fig. \ref{fig:6} that
the enhancement of the energy absorption is realized in the broader
range of the external field $B_{\rm ext}$ with the increase of the
normalized laser intensity $a_0$.
For a given $a_0$, we have performed a number of simulations with different
$B_{\rm ext}$ (e.g., Fig. \ref{fig:2}) and have evaluated the upper and lower
values of $B_{\rm ext}$ between which the absorption is above a
threshold defined as $\Delta \mathcal{E}_{\rm kin} / \mathcal{E}_0 = 0.1$.
The numerically obtained $B_{\rm upper}$ traces the prediction curve
given by Eq. (\ref{eq:bres}) with a small scattering by factor of
about 2.  
Note that the dependence of the upper limits on $a_0$ is unaffected
qualitatively by the laser polarization and ion mobility, as
demonstrated in Fig. \ref{fig:6}. 
It might be better to estimate $a_0$ in this figure from the amplitude
of the transmitted whistler wave, which becomes lower than that of the
incident laser.
But here, the difference is neglected for simplicity.

When the laser intensity is non-relativistic, $a_0 \lesssim 1$, the
upper limits has little dependence on $a_0$, and takes a constant
value $B_{\rm res}$ which is influenced by the Doppler shift.
If the intensity becomes relativistic, $a_0 \gtrsim 1$, the upper
limits increase in proportion to $a_0$ because of the relativistic
effects. 
On the other hand, the lower limits are always around $B_c$, unless
the relativistic transparency comes into play as seen at $a_0 \sim 100$
\cite{palaniyappan12,stark16}.
Thus the large conversion efficiency is induced at the broadened
conditions of $B_{\rm ext}$, which is shown by the gray region in
Fig. \ref{fig:6}. 

\begin{figure*}
\includegraphics[scale=1.0,clip]{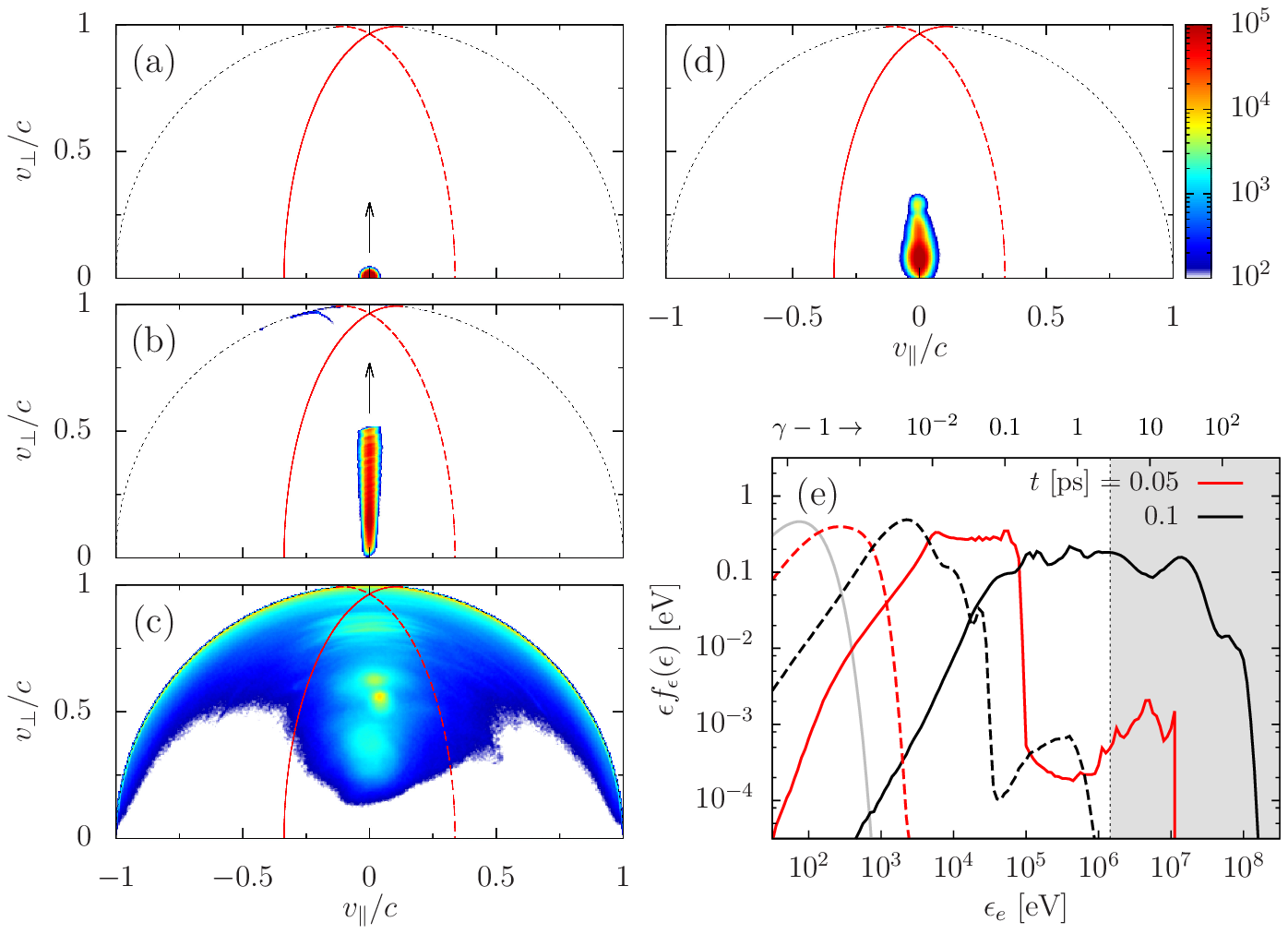}%
\caption{
(Color online)
(a-c)
Parallel and perpendicular velocity diagram of electrons,
$v_{\parallel}$-$v_{\perp}$, for the fiducial model observed at (a) $t
= 0$, (b) 0.05, and (c) 0.1 ps.   
The color denotes the particle number.
The cyclotron resonance condition given by Eq. (\ref{eq:beta}) is
shown by the solid and dashed curves. 
The dotted curve denotes the upper limit of
$|\mbox{\boldmath{$v$}}| = c$.
(d)
The same figure as panel (c) for a lower intensity $I_0 = 10^{19}$
W/cm$^2$. 
(e) 
Energy spectrum of electrons $\epsilon f_{\epsilon} (\epsilon)$ in the
fiducial (solid) and lower 
intensity models (dashed) taken at $t = 0.05$ (red) and 0.1 ps
(black). 
The gray curve is the initial Maxwellian spectrum of $T_e = 0.05$ keV.
The resonance condition $\gamma \gtrsim \gamma_{\rm res}$ is indicated
by the gray zone. 
\label{fig:3}}
\end{figure*}

\begin{figure*}
\includegraphics[scale=1.0,clip]{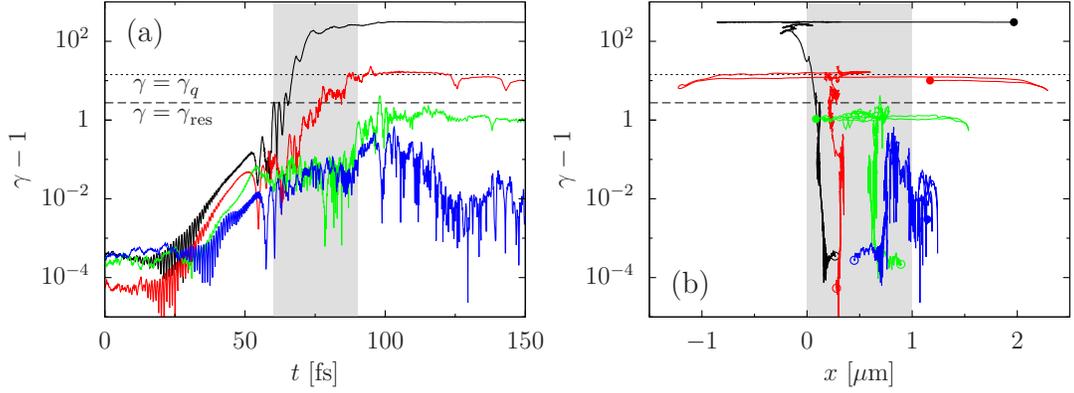}%
\caption{
(Color online)
Trajectories of four representative electrons in the fiducial model shown
by (a) the time-energy and (b) position-energy diagrams.
The same color in the both panels means the track of an identical
particle, which is drawn form $t = 0$ [open circles in panel (b)] to 150 fs
(filled circles).
Gray areas denote (a) the duration of 30 fs when the laser pulse is hitting
the foil target and (b) the initial target location of 1 $\mu$m thickness.
As a reference, the resonance condition $\gamma = \gamma_{\rm res}$
and the quiver energy $\gamma = \gamma_q$ are depicted by the
horizontal dashed and dotted lines, respectively.
\label{fig:7}}
\end{figure*}

Interestingly, the resonance process can be clearly identified in the
velocity diagram of electrons. 
The resonance condition is expressed by using the parallel and
perpendicular velocities, instead of $\gamma$, as 
\begin{equation}
\beta_{\perp}^2 = 1 - \beta_{\parallel}^2 -
\frac1{\tilde{B}_{\rm ext}^2}
\left( 1 -  \frac{\beta_{\parallel}}{\beta_{\phi}} 
\right)^2 \;.
\label{eq:beta}
\end{equation}
This condition for $v_{\parallel}$ and $v_{\perp}$ appears in the
velocity diagram as two lines shown in Fig. \ref{fig:3} for
the fiducial condition with $\tilde{B}_{\rm ext} = 3.74$.  
Since the external field is much larger than $B_c$, there exists no
resonant electrons at the beginning [Fig. \ref{fig:3}(a)]. 

However, due to the penetration of a whistler wave, the resonance is 
triggered by relativistic electrons accelerated by the laser's electric
field.
During the laser interaction, only the perpendicular velocity
increases toward the relativistic regime, while $v_{\parallel}$ is
almost unchanged and keeps the thermal distribution
[Fig. \ref{fig:3}(b)]. 
When $v_{\perp}$ hits the resonance condition at $\gamma \sim
\gamma_{\rm res}$ and $v_{\parallel} \sim 0$, further jump up of the
electron energy starts through the cyclotron resonance. 
It is this resonance point indeed that is used in the derivation of
$\gamma_{\rm res}$ [Eq. (\ref{eq:gres})].  

The electron temperature is highly anisotropic at this phase
($T_{\perp} \gg T_{\parallel}$), so that the
kinetic mirror instability could be driven \cite{hasegawa75,melrose86},
resulting in the thermalization of the electrons.
Then finally, high energy conversion is accomplished at the end of
calculation [Fig. \ref{fig:3}(c)].  

By contrast, the resonant acceleration cannot happen if the laser
intensity is lower.
For a model with the intensity $I_0 = 10^{19}$ W/cm$^2$, the
acceleration by the laser field is not enough to reach the resonance
point because $\gamma_q < \gamma_{\rm res}$ [Fig. \ref{fig:3}(d)].  
That is why there is no enhancement in the absorption efficiency.

Figure \ref{fig:3}(e) shows the probability density function for the
electron kinetic energy $\epsilon_e$ in the logarithmic binning
$\epsilon f_{\epsilon} (\epsilon) = f_{\ln \epsilon} (\ln \epsilon)$. 
For the fiducial model, the quiver kinetic energy, $\epsilon_q =
(\gamma_q - 1) m_e c^2$, can reach the requirement 
for the resonance $\gamma_{\rm res} - 1 \approx 2.7$ or $\epsilon_e
\approx 1.4$ MeV around the timing of $t = 0.05$ ps.    
Immediately after that, the maximum energy increases suddenly over 100
MeV by the resonant acceleration, which is much larger than the
free-electron ponderomotive limit $\epsilon_p = m_e c^2 a_0^2 / 2
\approx 60$ MeV \cite{sorokovikova16}. 
This could be happen only when the laser intensity satisfies
$a_0 \gtrsim \gamma_{\rm res}$.
The electron heating in the $I_0 = 10^{19}$ W/cm$^2$ run is not by
much and the energy of hot-electron component is at most
the quiver kinetic energy $\epsilon_q \approx 0.93$ MeV. 

Now, we will carefully inspect the trajectories of some representative
electrons in the fiducial run throughout the laser plasma interaction 
from $t =$ 0 to 150 fs.
Figure \ref{fig:7}(a) shows the time evolution of the kinetic energies
of selected four particles. 
When the transmitted laser light is propagating inside of the foil
as the whistler mode, the electrons gain the energy gradually. 
The perpendicular velocity is accelerated predominantly by the
CP electric field of the whistler wave.
The acceleration up to the quiver energy $\gamma = \gamma_q$
takes place at almost the same location, which corresponds to the
upward track in the position-energy diagram of Fig. \ref{fig:7}(b). 

When the peak energy of an electron is below the resonance condition
$\gamma = \gamma_{\rm res}$, it loses the energy finally after the
laser pulse is finished.
However, if an electron successfully gains the energy reaching
$\gamma = \gamma_{\rm res}$, it can be kicked by the cyclotron resonance.
Such electrons retain the gained energy (e.g., the 'red' trajectory in
Fig. \ref{fig:7}) and contribute the net absorption of the laser energy. 

It is found that the acceleration beyond the free-electron
ponderomotive limit $\epsilon_p$ commonly happens at just outside of
the front surface of the target. 
All the superponderomotive electrons follow the exactly same track in
Fig. \ref{fig:7}(b) when the energy passes far over the quiver energy
$\gamma = \gamma_q$.  
The incident laser is partly reflected at the surface, so that a
CP standing wave is formed there.
Direct acceleration by the static electric field of the standing
wave would be the mechanism to generate such extreme electrons.

\begin{figure}
\includegraphics[scale=1.0,clip]{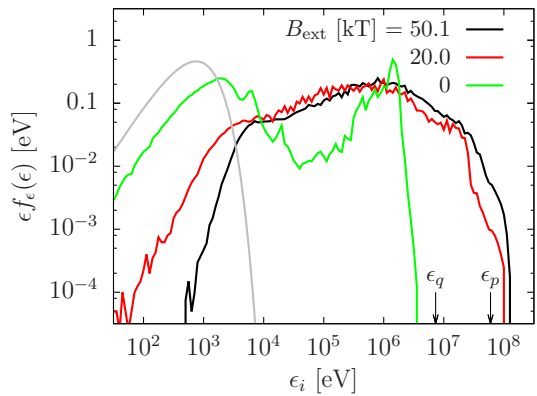}%
\caption{
(Color online)
Energy spectrum of ions $\epsilon f_{\epsilon} (\epsilon)$ in the
fiducial model (black) as well as the
different $B_{\rm ext}$ runs with 20.0 kT (red) and nothing (green).
The RCP laser with $I_0 = 10^{21}$ W/cm$^2$ is used in these models, and
all the spectra are taken at $t = 1$ ps.
The gray curve is the initial Maxwellian spectrum of $T_e = 0.5$ keV.
The quiver kinetic energy of electrons $\epsilon_q$ and the
free-electron ponderomotive limit $\epsilon_p$ are indicated by arrows. 
\label{fig:8}}
\end{figure}

\begin{figure*}
\includegraphics[scale=1.0,clip]{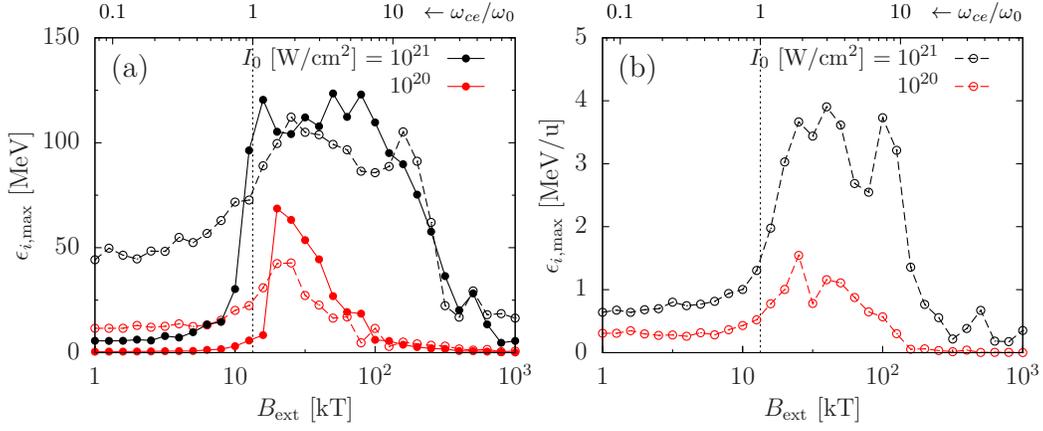}%
\caption{
(Color online)
(a) 
Dependence of the external field strength $B_{\rm ext}$ on the maximum
energy of ions $\epsilon_{i,\max}$ in the models of $I_0 = 10^{21}$
(black) and 10$^{20}$ W/cm$^2$ (red). 
The LP laser cases are also shown by the open circles.
The other parameters are identical to the models shown in Fig. \ref{fig:2}.
The dotted line indicates the critical field strength $\tilde{B}_{\rm
  ext} = 1$.
(b)
The maximum ion energy for the models of a fully ionized carbon
target. 
The LP laser is used and the other parameters are the identical to the
fiducial hydrogen model.   
\label{fig:4}}
\end{figure*}

\begin{figure*}
\includegraphics[scale=1.0,clip]{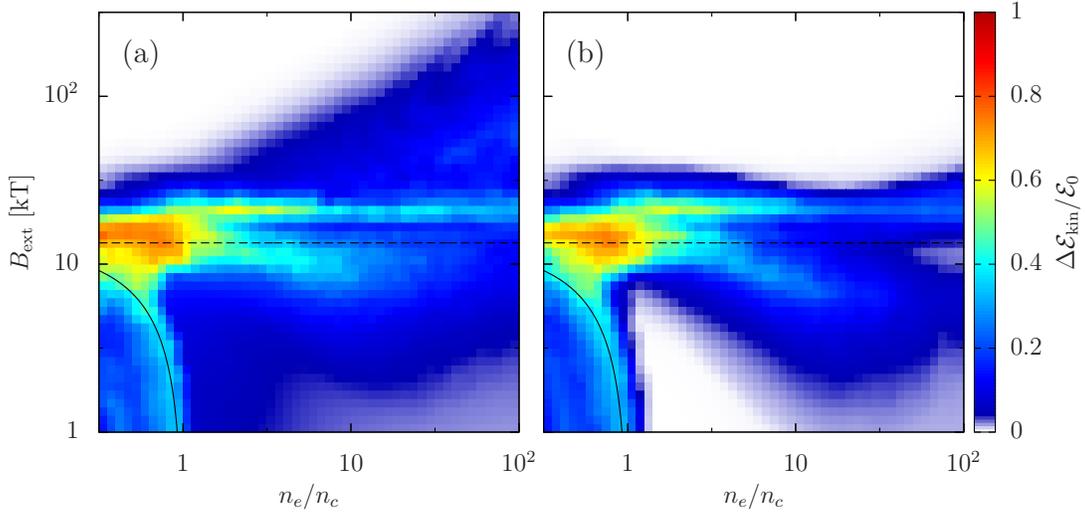}%
\caption{
(Color online)
(a) 
Energy absorption in the density and field strength diagram
constructed from the results of $50 \times 50$ simulations.  
The thickness of the hydrogen target is assumed to be 3 $\mu$m for
these calculations. 
The RCP laser is used with the intensity $I_0 = 10^{19}$
W/cm$^2$. 
The solid and dashed lines denote the R cutoff ($\tilde{n}_e =
1 - \tilde{B}_{\rm ext}$) and the cyclotron resonance ($\tilde{B}_{\rm
  ext} = 1$), respectively. 
(b)
The same figure as panel (a) for the immobile ion simulations.
\label{fig:5}}
\end{figure*}

\section{Discussion and Conclusions \label{sec:4}}

In summary, we have derived the relativistic condition of the
cyclotron resonance for the interaction of overdense plasmas with an
intense laser under a strong magnetic field.
The condition is benchmarked successfully by 1D PIC simulations.   
The energy absorption increases dramatically when the external
field strength is nearly equal to the critical value $B_c$.
Furthermore, the cyclotron resonance operates at a wider parameter range
when the laser is highly relativistic, $a_0 \gg 1$.
In the models of a stronger field than $B_c$, the cyclotron resonance
is allowed for the relativistic electrons only. 
On the other hand, the laser light propagates into overdense
plasmas without cutoff for that cases. 
With a help of the intense laser field, such relativistic electrons
are injected within the magnetized dense plasmas, and they trigger the
efficient plasma heating via the broadened cyclotron resonance. 

Ion acceleration is one of the potential applications
\cite{daido12,macchi13}. 
In this study, a large sheath potential is generated by hot electrons
at the both target surfaces. 
The maximum ion energy increases through the target normal sheath
acceleration (TNSA) \cite{wilks01} up to the comparable order of the
superponderomotive electrons, $\epsilon_{i, \max} \sim 100$ MeV.  
Since the TNSA is regarded as an energy conversion process from
electrons to ions, the maximum ion energy could have some correlation
with the electron energy \cite{sentoku03}.
Figure \ref{fig:8} shows the energy spectrum of ions for various
$B_{\rm ext}$ cases.
For the case of $B_{\rm ext} = 0$, the maximum ion energy is of the
order of the quiver kinetic energy of electrons, that is $\epsilon_q
\approx$ 7.3 MeV for this model.
When the energy absorption by electrons is largely enhanced by the
presence of a strong magnetic field, 
a significant fraction of ions is actually accelerated above the
maximum energy in the unmagnetized run.

A good correlation between the maximum ion energy and the conversion
efficiency can be seen in Fig. \ref{fig:4}(a).
This is also true for the LP laser cases, unless the
$\mbox{\boldmath{$J$}} \times \mbox{\boldmath{$B$}}$ force becomes
dominant at $B_{\rm ext} \lesssim B_c$.    
For medical applications, heavy ion acceleration would be an
interesting topic.
When a 1 $\mu$m-thick carbon target with $\tilde{n}_e = 603$
(equivalent to the mass density of diamond $\rho = 3.51$ g/cm$^3$) is
irradiated by a LP laser with $I_0 = 10^{21}$ W/cm$^2$, the maximum energy
is about 4 MeV/u at the range of $\tilde{B}_{\rm ext} \sim 1$-10
[Fig. \ref{fig:4}(b)]. 
Here, the wide resonance range of $B_{\rm ext}$ is the great advantage
for the practical verification of this mechanism in future.   

Direct plasma heating by electromagnetic waves might be attractive as
an alternative scheme for the ICF.
The systematic study of the propagation properties of whistler waves
in dense plasmas should be a quite important next step.
It is informative to extend the parameter space of the density and
field strength, and make a chart like the Clemmow-Mullaly-Allis (CMA)
diagram for the laser plasma interaction. 
Figure \ref{fig:5} is an example for the cases of the RCP laser with
$I_0 = 10^{19}$ W/cm$^2$. 
The spots of high absorption trace the lines of the R cutoff and
cyclotron resonance. 
Obviously the behavior of underdense plasmas is also affected by the
external field, especially around $B_{\rm ext} \sim B_c$.
When $B_{\rm ext} < B_c$, the laser light cannot enter the target if
the density is higher than the R cutoff, and then it is mostly
reflected at the surface. 

In principle, the whistler waves can propagate into any density if
$B_{\rm ext} > B_c$, and deliver the energy directly to dense plasmas
without going through hot electrons.
This feature suggests a totally different way of the use of strong
magnetic fields in the ICF plasmas from the previous work
\cite{hohenberger12,perkins13,wang15,fujioka16}.
However the high absorption becomes an obstacle in terms of the
propagation.
In other words, the laser transmittance will be reduced by the amount
of the energy absorption.
The upper half of Fig. \ref{fig:5}(a), where $B_{\rm ext} > B_c$,
shows that the energy absorption is larger when the density $n_e$ is
larger and $B_{\rm ext}$ is closer to $B_c$. 
But this feature disappears cleanly if the ion motion is
inhibited, meaning that ion acoustic waves driven by the Brillouin
instability \cite{forslund72} would play an important role for the
high energy conversion there. 
The direct ion heating by whistler waves will be discussed in detail
in the subsequent paper.    


\begin{acknowledgments}
We thank Y. Sakawa, S. Fujioka, A. Arefiev, and D. Kawahito for
fruitful discussions and M. Zosa for his careful reading of the
manuscript.   
We also thank anonymous referees for their constructive comments.
This work was supported by JSPS KAKENHI Grant Numbers JP26287147,
JP15K21767, and JP16H02245. 
Computations were carried out on HCC at the Cybermedia Center and
SX-ACE at the Institute of Laser Engineering of Osaka University. 
\end{acknowledgments}

%





\end{document}